\documentclass[useAMS,usenatbib]{mn2e}
\usepackage{graphicx}
\usepackage{subfigure}

\newcommand{\Mjup}{\mbox{M$_{\rm Jup}$}}

\newcommand{\Msun}{\mbox{M$_{\odot}$}}
\newcommand{\degrees}{$^{\rm{o}}$}
\newcommand{\ms}{\mbox{m\,s$^{-1}$}}
\newcommand{\kms}{\mbox{km \ s$^{-1}$}}
\newcommand{\Rsun}{\mbox{R$_{\odot}$}}

\newcommand{\Lsun}{\mbox{L$_{\odot}$}}

\newcommand{\aap}{A\&A}
\newcommand{\apj}{ApJ}

\newcommand{\apjs}{ApJS}
\newcommand{\pasp}{PASP}
\newcommand{\mnras}{MNRAS}
\newcommand{\pasj}{PASJ}
\newcommand{\aj}{AJ}
\newcommand{\ltsimeq}{\raisebox{-0.6ex}{$\,\stackrel
         {\raisebox{-.2ex}{$\textstyle <$}}{\sim}\,$}}
\newcommand{\gtsimeq}{\raisebox{-0.6ex}{$\,\stackrel
         {\raisebox{-.2ex}{$\textstyle >$}}{\sim}\,$}}

\title[The Pan-Pacific Planet Search III: Five companions orbiting giant stars]{The 
Pan-Pacific Planet Search III: Five companions orbiting giant stars}

\author[R.A. Wittenmyer, et al.]{R.A. Wittenmyer$^{1}$\thanks{E-mail: 
rob@unsw.edu.au (RW)}, R.P. Butler$^{2}$, L. Wang$^{3}$, C. 
Bergmann$^{4}$, G.S. Salter$^{5}$, \newauthor C.G. Tinney$^{1}$, and J.A. 
Johnson$^{6}$\\
$^{1}$School of Physics, University of New South Wales, Sydney 2052, Australia\\
$^{2}$Department of Terrestrial Magnetism, Carnegie Institution of Washington, 5241 Broad Branch Road, NW, Washington, DC 20015-1305, USA\\ 
$^{3}$Key Laboratory of Optical Astronomy, National Astronomical Observatories, Chinese Academy of Sciences, A20 Datun Road, Chaoyang District, Beijing 100012, China\\ 
$^{4}$University of Canterbury, Department of Physics and Astronomy, Christchurch 8041, New Zealand\\
$^{5}$Aix Marseille Universit\'e, CNRS, LAM (Laboratoire d'Astrophysique de Marseille) UMR 7326, 13388, Marseille, France\\
$^{6}$Harvard-Smithsonian Center for Astrophysics, Cambridge, MA 02138 USA}

\begin{document}

\date{Accepted  Received ; in original form }

\pagerange{\pageref{firstpage}--\pageref{lastpage}} \pubyear{2015}

\maketitle

\label{firstpage}

\begin{abstract}

We report a new giant planet orbiting the K giant HD\,155233, as well as 
four stellar-mass companions from the Pan-Pacific Planet Search, a 
southern hemisphere radial velocity survey for planets orbiting nearby 
giants and subgiants.  We also present updated velocities and a refined 
orbit for HD\,47205b (7 CMa b), the first planet discovered by this 
survey.  HD\,155233b has a period of 885$\pm$63 days, eccentricity 
$e=0.03\pm$0.20, and m~sin~$i$=2.0$\pm$0.5\,\Mjup.  The stellar-mass 
companions range in m~sin~$i$ from 0.066\,\Msun\ to 0.33\,\Msun.  Whilst 
HD\,104358B falls slightly below the traditional 0.08\,\Msun\ 
hydrogen-burning mass limit, and is hence a brown dwarf candidate, we 
estimate only a 50\% \textit{a priori} probability of a truly substellar 
mass.

\end{abstract}

\begin{keywords}

planetary systems, stars: giants, techniques: radial velocity

\end{keywords}

\section{Introduction}

Radial velocity planet search efforts have been underway for more than 
20 years, discovering hundreds of new planetary companions.  However, 
brown-dwarf and stellar-mass companions tend to be largely ignored in 
these surveys: limited telescope time means that targets showing 
large-amplitude variations ($\gtsimeq$500\,\ms) are usually dropped from 
the observing queue.  Only limited results on such massive companions 
have been published \citep{patel07}, with many objects featuring 
incomplete orbits.  However, such objects remain valuable for exploring 
the lower end of the mass function and the properties of stellar systems 
in the Solar neighbourhood.  The observed deficit of companions between 
13-80\,\Mjup\ is known as the ``brown dwarf desert'' \citep{marcy00, 
mazeh03}, and became evident in the earliest days of radial velocity 
planet searches \citep{campbell88, murdoch93}.  A comprehensive, 
self-consistent study by \citet{gl06} confirmed the presence of a 
``valley'' in the brown-dwarf mass range at $M=31^{+25}_{-18}$ \Mjup.  
Stellar companions are also not to be ignored, since any binary system 
with a well-characterised orbit is useful for a range of follow-up 
science.  Most obviously, a precisely-determined binary orbit permits 
the search for additional Doppler velocity signals due to planets 
orbiting one or both stars.  Binary systems compose $\sim$50\% of nearby 
star systems \citep{dm91, rag10}, yet are hosts to only $\sim$5\% of 
known extrasolar planets.  The mechanisms and outcomes of planet 
formation in close binary systems remain significant questions virtually 
unexplored by observations.

At the 3.9m Anglo-Australian Telescope (AAT), we carried out a 5-year 
survey for planets orbiting intermediate-mass evolved stars in an effort 
to characterise the dependence of planetary system properties on stellar 
mass \citep{47205paper, reffert15}.  \textbf{The initial sample was 
selected accoring to these criteria: $1.0 < (B-V) < 1.2$, $1.8 < M_V < 
3.0$, and $V<8.0$.  Of those stars meeting these criteria, 18 were 
discarded as binaries by their \textit{Hipparcos} multiple systems 
flag.}  From the 167-star Pan-Pacific Planet Search (PPPS) sample, 17 
targets turned out to be double-lined spectroscopic binaries (SB2s -- 
Table~\ref{sb2}).  Since the Doppler velocity technique used here 
assumes only a single set of spectral lines, we were unable to derive 
velocities for these binaries and they were dropped from the program as 
soon as they were identified as such.  We note that it has recently 
become possible to modify planet-search Doppler codes to obtain precise 
velocities for SB2s, by including the secondary star's spectrum in the 
modelling process if the flux ratio is known.  This novel approach is 
now being applied to the Alpha Centauri binary system 
\citep{bergmann15,endl15}.

A further 31 stars in the PPPS sample show large-amplitude velocity 
variations or long-term trends, which indicate stellar-mass companions.  
This yields a first-order binary fraction of 48/167=29\%, for a sample 
which was initially selected to avoid suspected binaries 
\citep{47205paper}.  As is common in planet-search programs, these stars 
were deprecated in observing priority and so the orbits of these massive 
companions cannot be constrained.  However, four stars host 
shorter-period companions and received enough observations to reliably 
determine the orbital solutions.  Here, we present the velocity data and 
orbital solutions for stellar-mass companions orbiting HD\,34851, 
HD\,94386, HD\,104358, and HD\,188981.  We also present new data and a 
refined orbital solution for the giant planet orbiting 7 CMa 
(HD\,47205).  Section~2 briefly describes the observational data and 
gives the stellar parameters.  In Section~3, we describe the 
orbit-fitting process.  Section 4 gives the companion parameters and our 
conclusions.

\section{AAT Observations and Stellar Properties}

The PPPS used the UCLES echelle spectrograph \citep{diego:90} at the AAT 
to obtain high-resolution spectra, with an observing procedure identical 
to that used by the 16-year Anglo-Australian Planet Search (AAPS).  
UCLES uses a 1-arcsecond slit to deliver a resolving power of 
$R\sim\,45,000$.  An iodine absorption cell, temperature-controlled at 
60.0$\pm$0.1$^{\rm{o}}$C, is placed in the light path.  The iodine cell 
imprints a forest of narrow absorption lines from 5000 to 6200\,\AA, 
allowing simultaneous calibration of instrumental drifts as well as a 
precise wavelength reference \citep{val:95,BuMaWi96}.  Precise Doppler 
velocities are derived using the well-established PSF modelling 
techniques described in \citet{BuMaWi96}.  The result is a velocity with 
a zero-point relative to the stellar template: a high S/N spectrum 
obtained without the iodine cell at $R\sim\,60,000$.  The internal 
uncertainty estimate includes the effects of photon-counting 
uncertainties, residual errors in the spectrograph PSF model, and 
variation in the underlying spectrum between the iodine-free template 
and epoch spectra observed through the iodine cell.  A summary of the 
observations is given in Table~\ref{obslog}, and the radial velocities 
are given in Tables~\ref{34851vels}--\ref{188981vels}.


For each new star discussed in this work, we have used our iodine-free 
template spectrum ($R\sim\,60,000$, S/N$\sim$150-250) to derive 
spectroscopic stellar parameters.  The same techniques were used in 
\citet{47205paper} and \citet{121056} for HD\,47205 and HD\,121056, 
respectively.  In brief, the iron abundance [Fe/H] was determined from 
the equivalent widths of $\sim$30 unblended Fe lines, and the LTE model 
atmospheres adopted in this work were interpolated from the ODFNEW grid 
of ATLAS9 \citep{Castelli2004}.  The effective temperature 
($T_\mathrm{eff}$) and bolometric correction ($BC$) were derived from 
the colour index $B-V$ and the estimated metallicity using the empirical 
calibration of \citet{Alonso1999,Alonso2001}.  Since the 
colour-$T_\mathrm{eff}$ method is not extinction-free, we corrected for 
reddening using $E(B-V)$ \citep{schlegel98}.  The stellar mass and age 
were estimated from the interpolation of Yonsei-Yale ($\mathrm{Y}^2$) 
stellar evolution tracks \citep{Yi2003}.  The resulting stellar masses 
were adopted for calculating the planet masses.  Complete stellar 
parameters from this analysis are given in 
Table~\ref{stellarparameters}.

\section{Orbit Fitting and Companion Parameters}

As shown in Table~\ref{obslog}, many of the stars considered here 
received fewer observations than is typical or desired for planet-search 
efforts.  However, the signals in our data are large enough as to be 
unambiguous, particularly for the stellar-mass companions.  We first 
used a genetic algorithm to search a wide range of orbital periods, 
running for 10,000 iterations (about $10^6$ possible configurations).  
This approach has been used in our previous work on systems for which 
data are sparse and/or poorly sampled \citep[e.g.][]{tinney11, 
47205paper,121056}.  In all cases, convergence occurred rapidly, a 
hallmark of a genuine signal.  Again as in our previous work, we then 
used the best solution from the genetic algorithm as a starting point 
for the generalized least-squares program \textit{GaussFit} 
\citep{jefferys88}, here used to solve a Keplerian radial velocity orbit 
model.  Finally, we estimated the parameter uncertainties using the 
bootstrap routine within \textit{Systemic 2} \citep{mes09} on 10,000 
synthetic data set realisations.  The results are given in 
Table~\ref{planetparams}.

\section{Results}

We have continued to observe HD\,47205 in the four years since the 
discovery of HD\,47205b \citep{47205paper}.  \textbf{We have added 6 new 
epochs, extending the baseline by 964 days, a factor of two longer than 
reported in the discovery work.}  The updated orbital solution given in 
Table~\ref{planetparams} is significantly more precise, and has an rms 
of only 6.5\,\ms, further strengthening the case for the planet's 
presence and apparent solitude.  \textbf{Notably, the orbital period is 
now slightly longer (796d compared to 763d); other parameters remain 
within 1$\sigma$ of their originally-reported values.}  No residual 
trends or periodicities are evident, though of course very low-mass 
planets could remain undetected with our radial velocity precision and 
cadence \citep[e.g.][]{wittenmyer09, swift15}.

\subsection{A giant planet orbiting HD\,155233 }

We have obtained 21 radial velocity measurements of HD\,155233, 
indicating a linear trend and a periodic signal, though with significant 
jitter.  \textbf{A simple linear fit to our data yields a residual rms 
scatter of 24.2\,\ms\ and $\chi^2_{nu}=5.23$.  By comparison, a 
planet-only model (no trend) gives an rms of 19.1\,\ms\ and 
$\chi^2_{nu}=3.65$.  We fit a planet+trend model, which indicates a 
planetary companion with $P=885\pm$63 days, $e=0.03\pm$0.20, and 
m~sin~$i=2.0\pm$0.5 \Mjup\ (Table~\ref{planetparams}).  That model has a 
residual rms of 11.0\,\ms\ and $\chi^2_{nu}=1.44$, with a linear trend 
of 15.4$\pm$6.5 m\,s$^{-1}$\,yr$^{-1}$.  \citet{feng15} gave a 
discussion on estimating minimum companion masses from velocity data 
with such trends when no curvature is evident.  Using their Equation 1, 
which assumes an ``unlucky observer'' seeing an orbital period 
$P\sim\,1.25T_{obs}$ (=4.9\,yr) and $e\sim\,0.5$, we obtain a minimum 
mass of 2.0\,\Mjup\ for the distant outer body.  The possibility also 
exists that the velocity signals of the planet and/or the long-term 
trend originate from a magnetic activity cycle; however, the velocity 
amplitudes so induced are less than 10\,\ms\ \citep[e.g.][]{robertson13, 
robertson15, fulton15}. }

\begin{figure}
\begin{tabular}{cc}
\includegraphics[width=80mm]{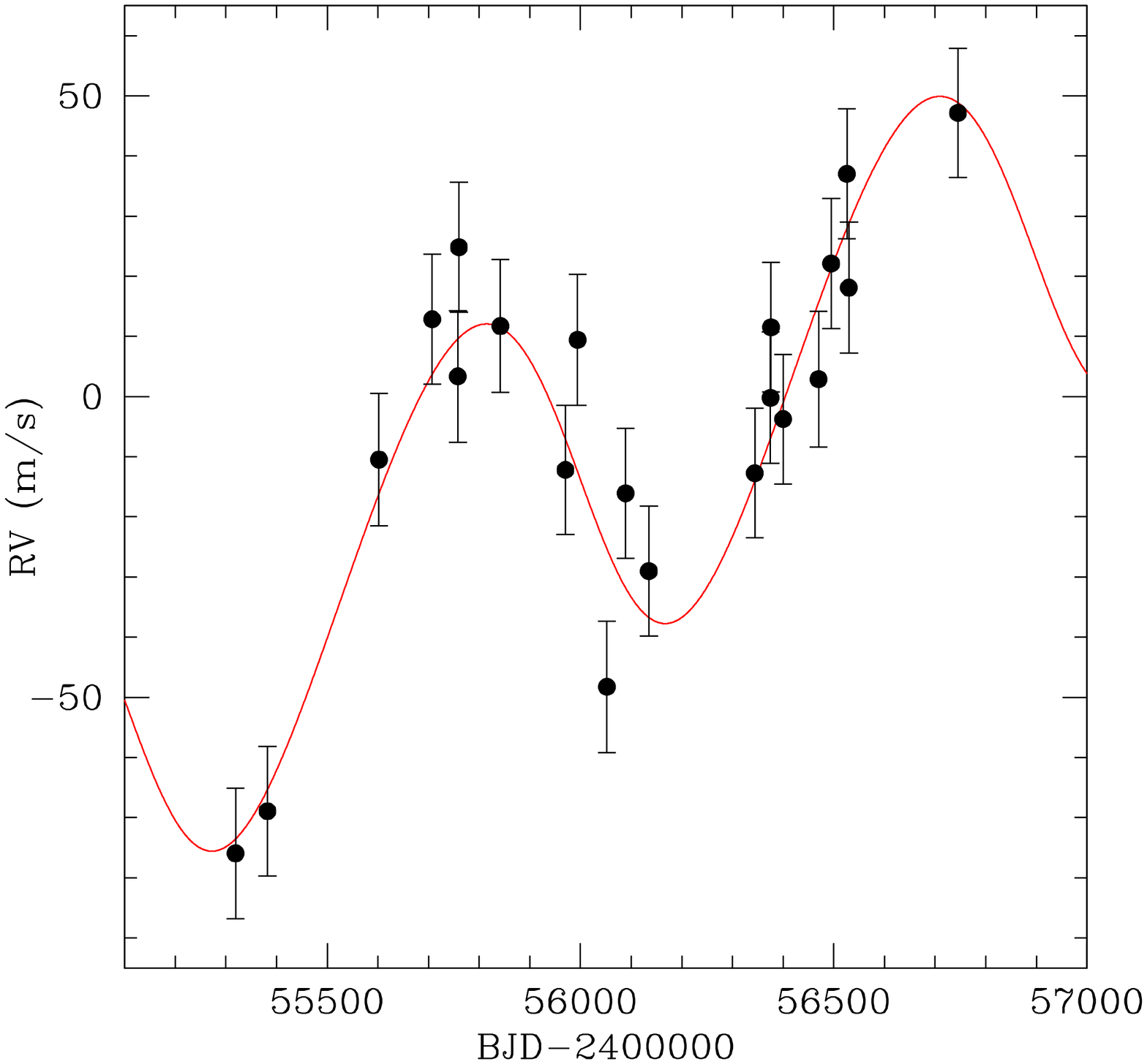} \\
\includegraphics[width=80mm]{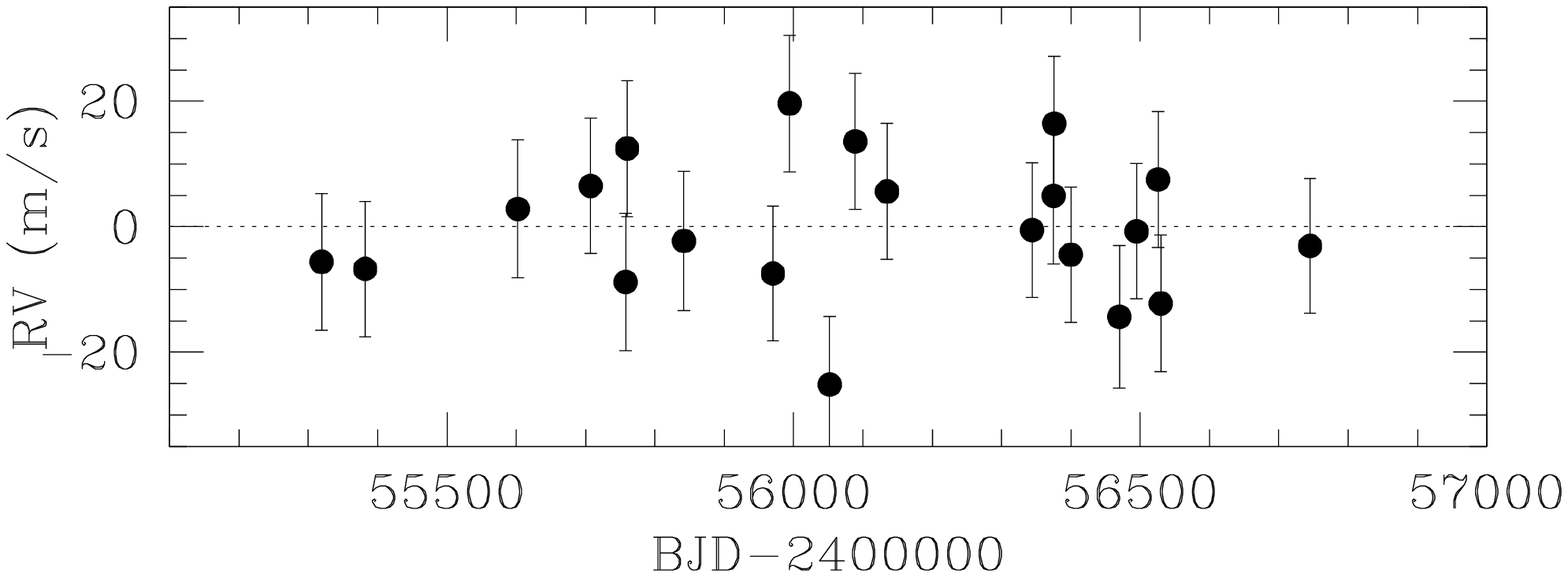} \\
\end{tabular}
\caption{Top panel: AAT data and Keplerian fit for a 2\Mjup\ planet 
orbiting HD\,155233.  A linear trend of 
15.4$\pm$6.5\,m\,s$^{-1}$\,yr$^{-1}$ is included in the fit as a free 
parameter and is shown.  Error bars include 10.65\,\ms\ of jitter added 
in quadrature. The rms about this fit is 10.9\,\ms; residuals are shown 
in the bottom panel. }
\label{155233}
\end{figure}

There remains substantial uncertainty in the fit due to the best-fit 
10.65\,\ms\ velocity jitter for the host star.  The residuals to the 
single planet + trend fit show no coherent periodicity, leading us to 
conclude that the residual scatter is best explained by stellar noise, 
as is common for giants \citep{hekker08,jones15}.  We also note that 
HD\,155233 has a relatively high projected rotational velocity $v$ sin 
$i=4.4\pm$1.0\kms; this is similar to HD\,34851 and HD\,104358, which 
have residual scatters of 34.6\,\ms\ and 15.3\,\ms, respectively 
(Table~\ref{planetparams}).  The data and model fit are shown in 
Figure~\ref{155233}.

\begin{figure}
\includegraphics[scale=0.4]{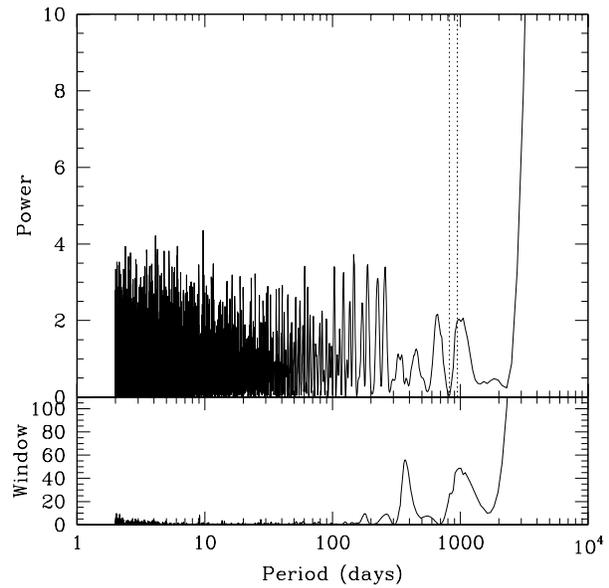}
\caption{Generalised Lomb-Scargle periodogram of ASAS photometry for 
HD\,155233.  A total of 218 epochs spanning 7.4 years yield no 
significant periodicities.  The $\pm\,1\sigma$ range of the planet's 
orbital period is shown as vertical dashed lines (885$\pm$63\,d). }
\label{155233pgram}
\end{figure}

To check whether the observed velocity variations could be due to 
intrinsic stellar processes, we examined the All-Sky Automated Survey 
(ASAS) $V$ band photometric data for HD\,155233 \citep{asas}.  A total 
of 222 epochs were obtained from the ASAS All Star 
Catalogue\footnote{http://www.astrouw.edu.pl/asas}.  After removing 
$5\sigma$ outliers, 218 points remained, and their generalised 
Lomb-Scargle periodogram \citep{zk09} is shown in 
Figure~\ref{155233pgram}.  No periodicities of interest are evident in 
data spanning 7.4 years.  Furthermore, we can use the projected 
rotational velocity v~sin~$i$ and the radius estimate from 
Table~\ref{stellarparameters} to estimate a maximum rotation period of 
58 days, well away from the planetary orbital period of 885 days.

\subsection{Four stellar-mass companions}

\begin{figure}
\begin{tabular}{cc}
\includegraphics[width=80mm]{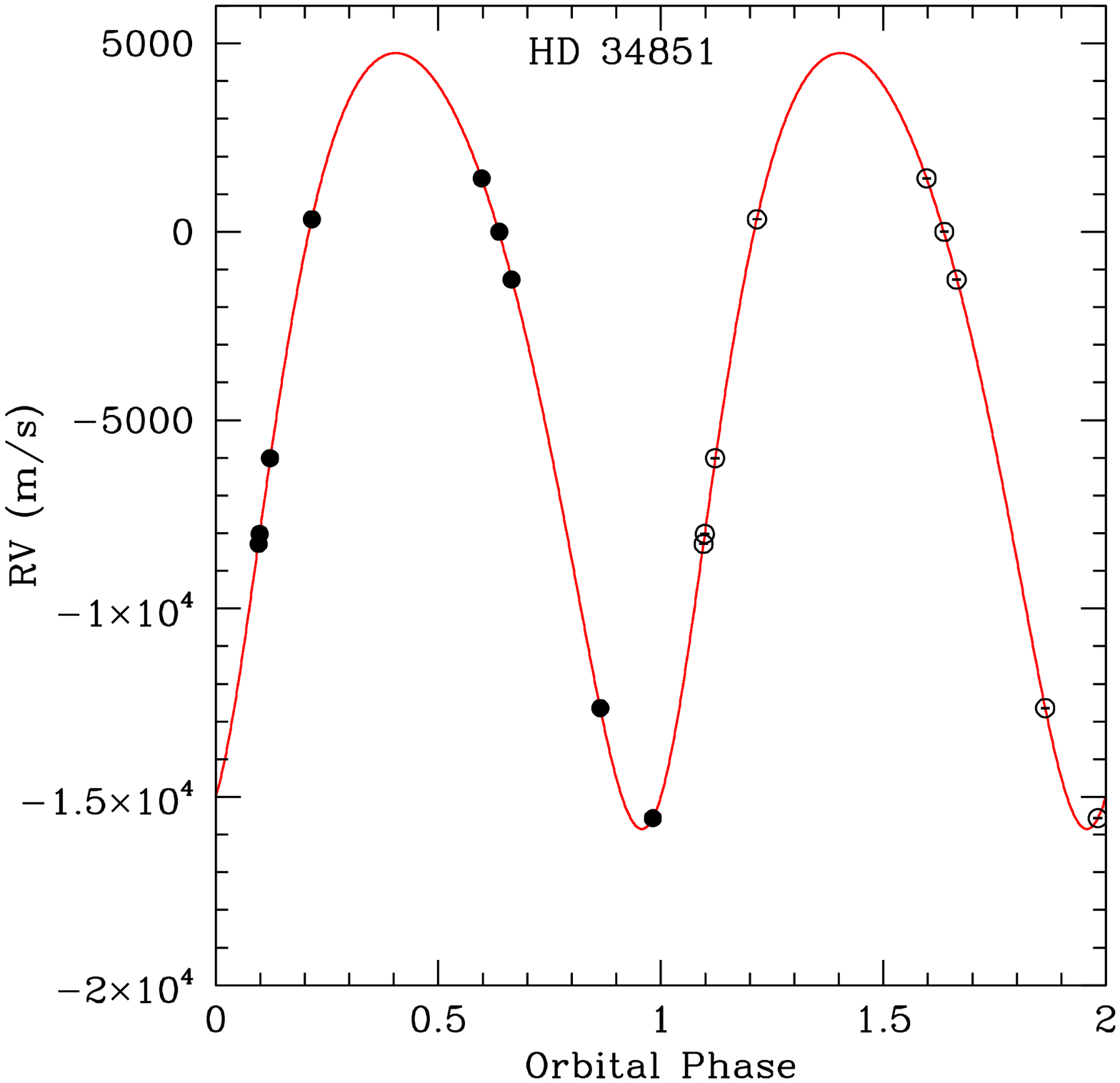} \\
\includegraphics[width=80mm]{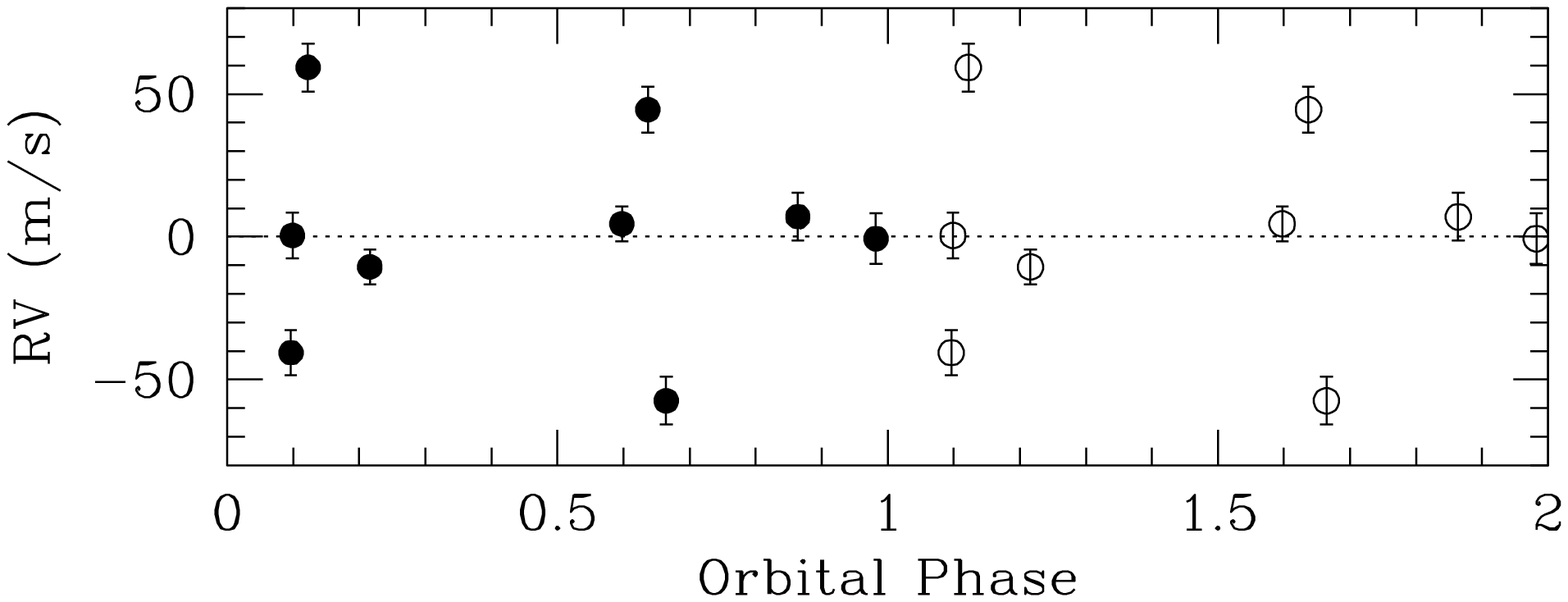} \\
\end{tabular}
\caption{AAT radial velocity data and Keplerian orbit fit for HD\,34851B 
(top) and residuals to the fit (bottom).  The data are phase-folded and 
two cycles are shown for clarity. }
 \label{binariesfits1}
\end{figure}

For HD\,34851, the rms about the fit is 34.6\,\ms, which is unusually 
large given that the typical velocity jitter of evolved stars targeted 
by the PPPS and other surveys is 5-20\ms\ 
\citep{sato05,johnson10,jones13}.  The companion has a minimum mass of 
345\,\Mjup, or 0.33\,\Msun, corresponding to a mid-M dwarf.  The 
Keplerian orbit fit for HD\,34851B and the other three stellar-mass 
companions are shown in Figure~\ref{binariesfits1}.

\begin{figure}
\begin{tabular}{cc}
\includegraphics[width=80mm]{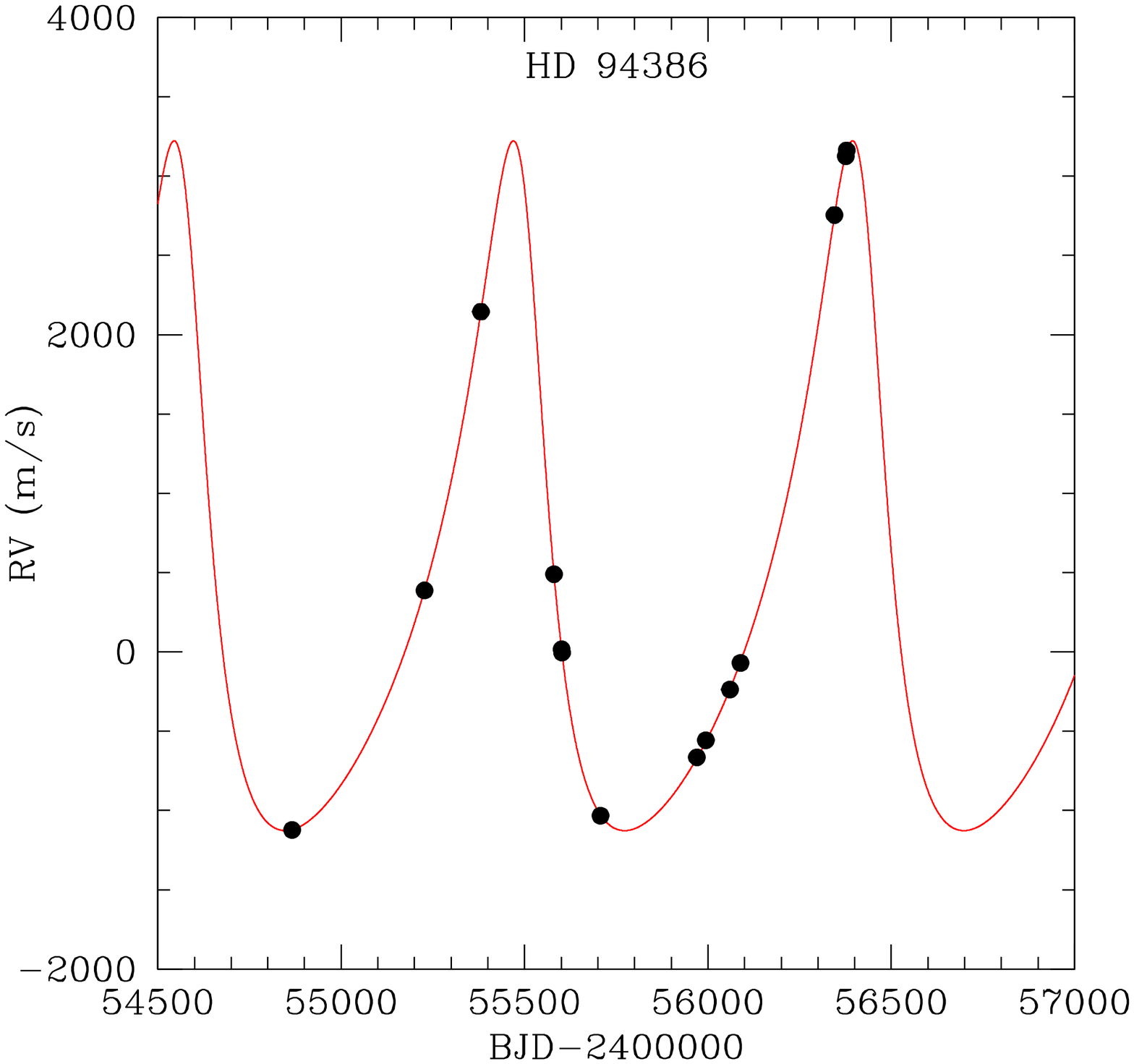} \\
\includegraphics[width=80mm]{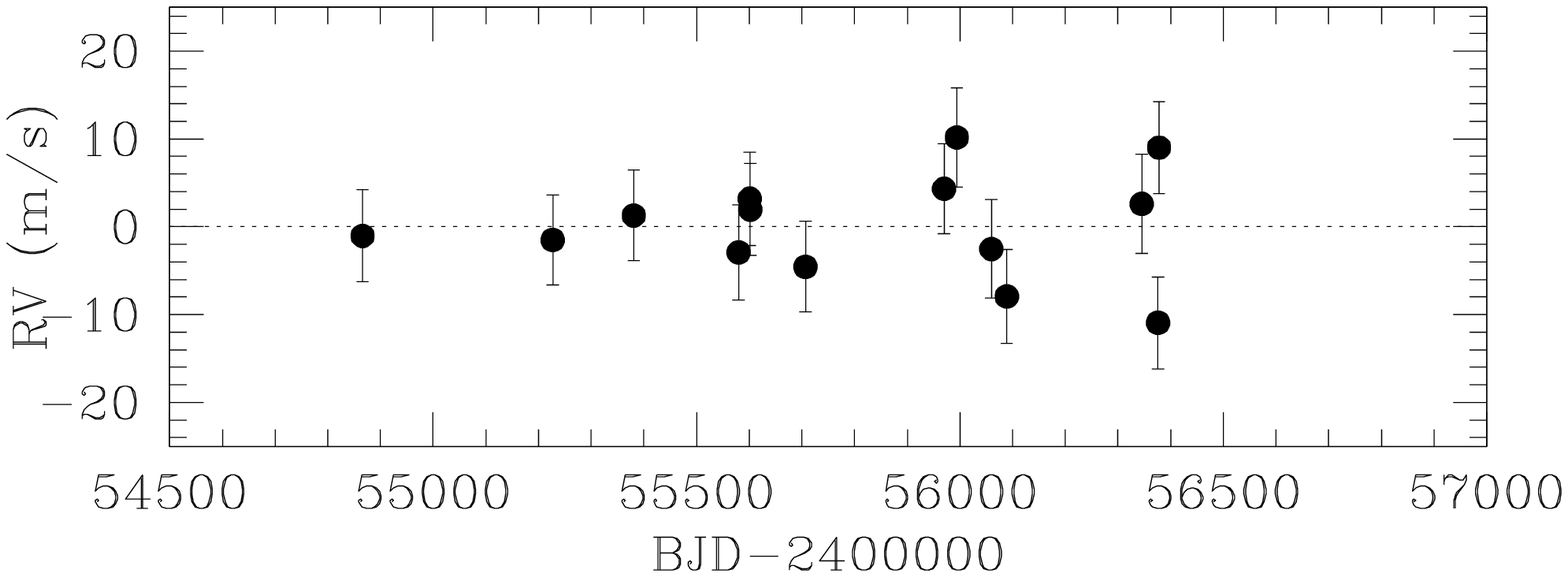} \\
\end{tabular}
\caption{AAT radial velocity data and Keplerian orbit fit for HD\,94386B 
(top) and residuals to the fit (bottom). }
 \label{binariesfits2}
\end{figure}
\begin{figure}
\begin{tabular}{cc}
\includegraphics[width=80mm]{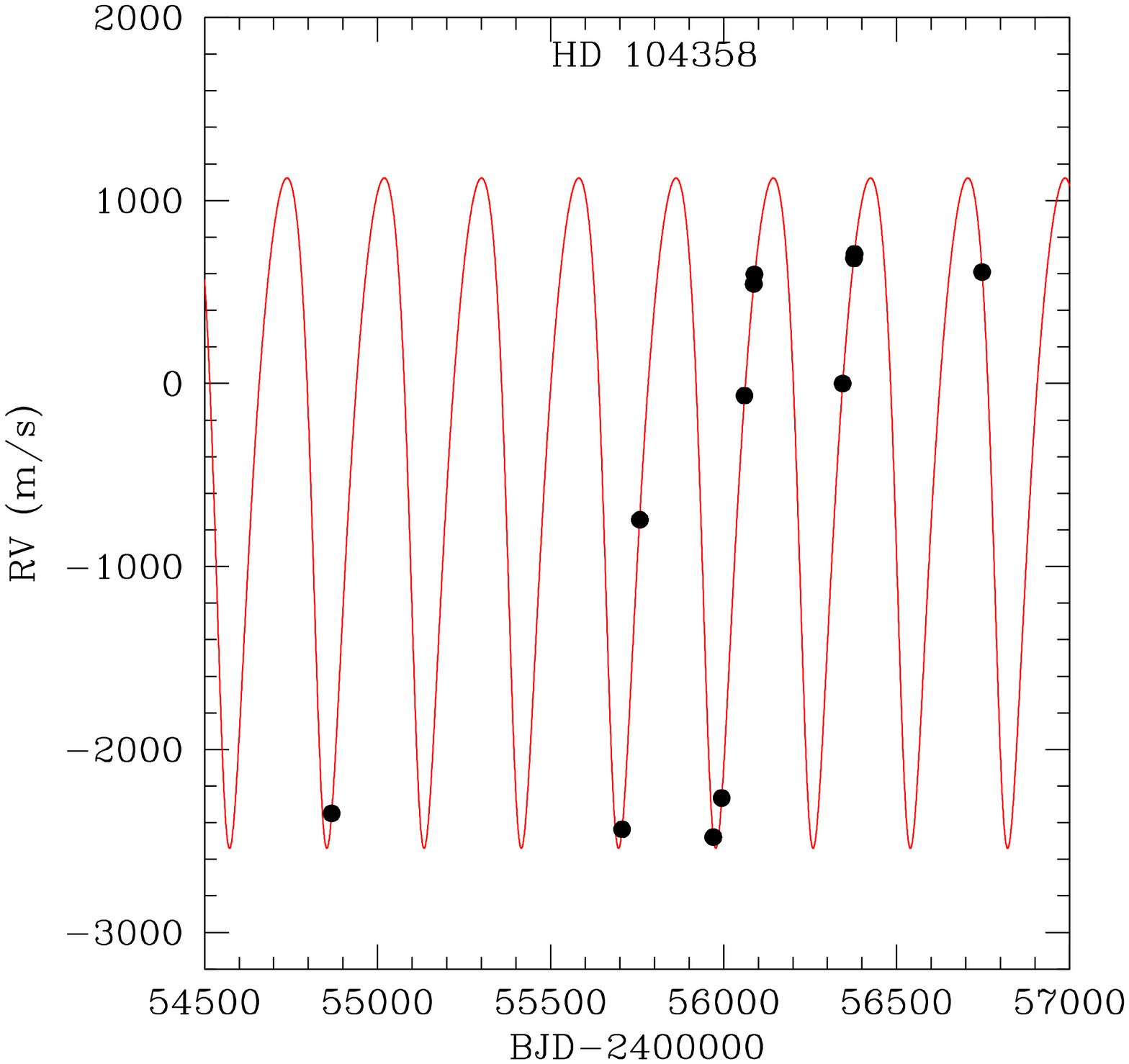} \\
\includegraphics[width=80mm]{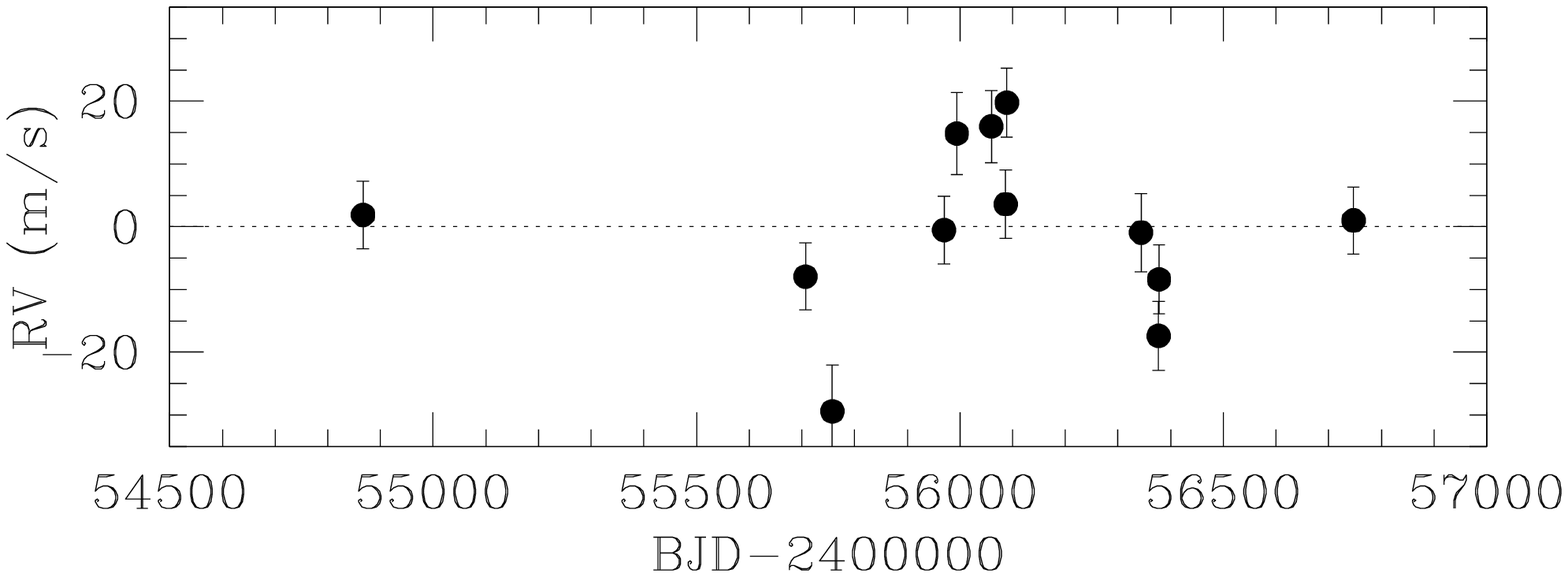} \\
\end{tabular}
\caption{AAT radial velocity data and Keplerian orbit fit for 
HD\,104358B (top) and residuals to the fit (bottom). }
 \label{binariesfits3}
\end{figure}

Nearly two orbital cycles of data result in a well-constrained fit for 
the massive companion orbiting HD\,94386, with an rms of only 5.7\,\ms\ 
(Figure~\ref{binariesfits2}).  The companion has m~sin~$i$ of 
112.5\,\Mjup\ ($=0.11$\,\Msun) corresponding to a late M dwarf.  
HD\,104358 hosts a brown dwarf candidate, with 
m~sin~$i=$69.3$\pm$1.9\,\Mjup\ ($=0.066$\,\Msun).  The probability of an 
inclination small enough to give a stellar-mass companion ($>$80\,\Mjup) 
is then 50\%.  The rms about the fit is 15.3\,\ms; removal of the 
4$\sigma$ outlier at BJD 2455757 gives the same parameters for the 
companion, though with slightly larger bootstrap uncertainties due to 
the already-limited amount of data available.  Hence we have retained 
all our data for the fitting.  The data and model fit are plotted in 
Figure~\ref{binariesfits3}.

\begin{figure}
\begin{tabular}{cc}
\includegraphics[width=80mm]{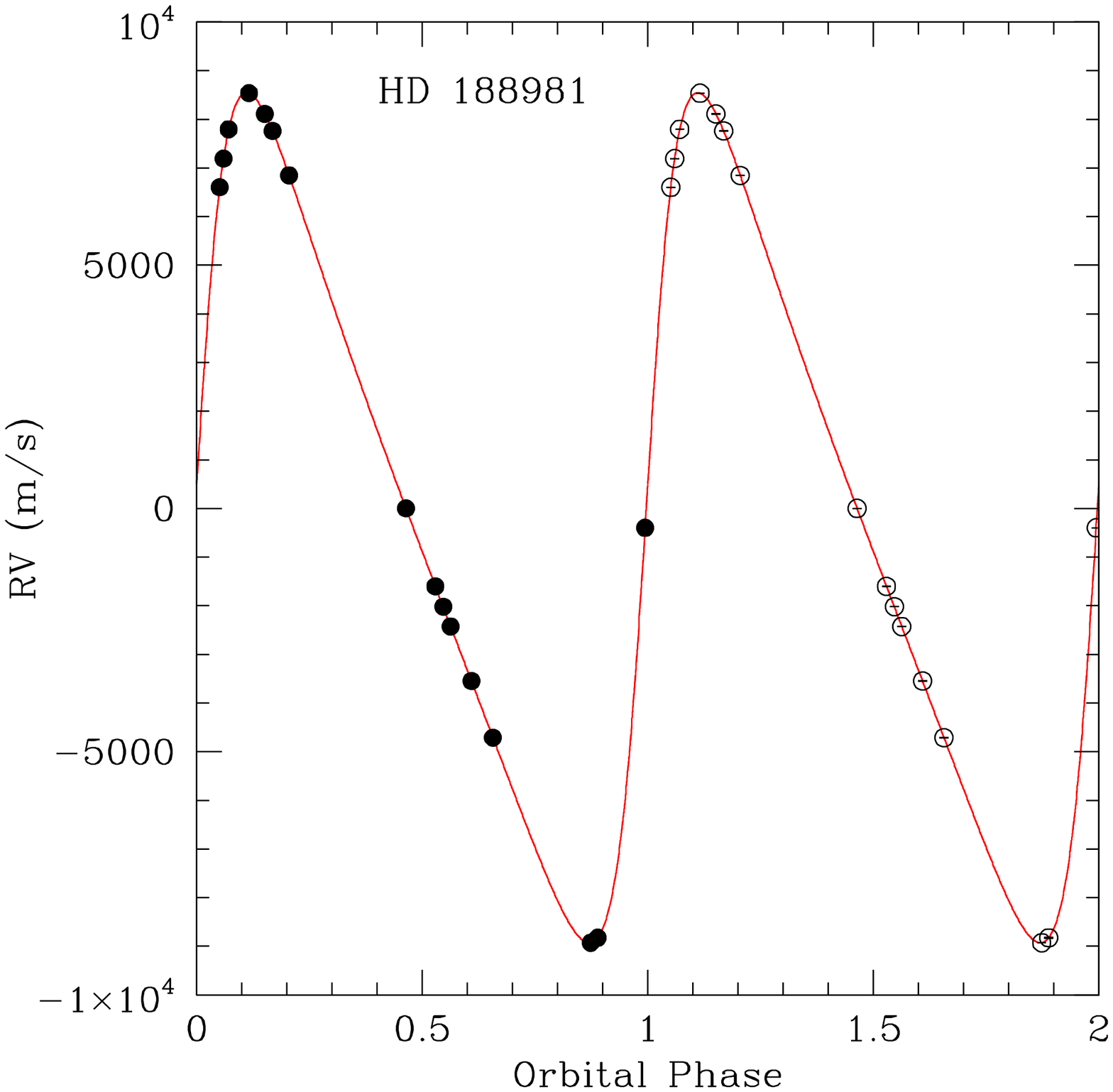} \\
\includegraphics[width=80mm]{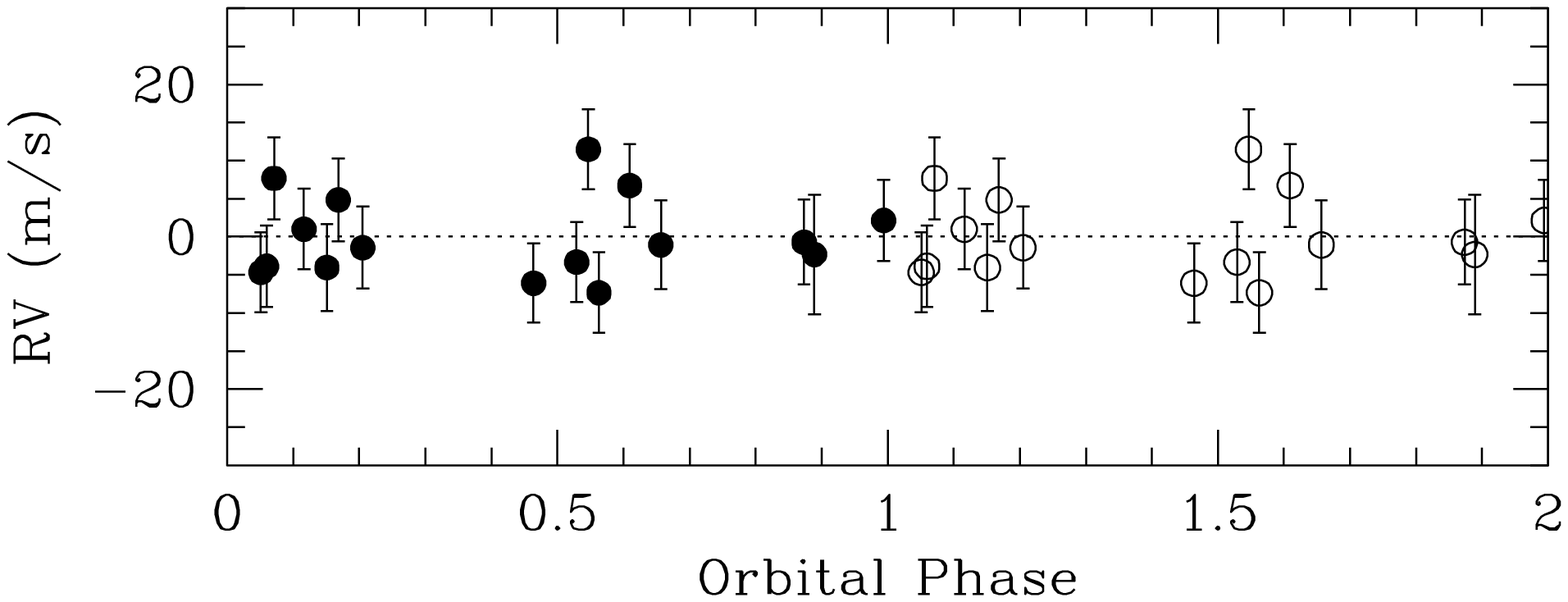} \\
\end{tabular}
\caption{AAT radial velocity data and Keplerian orbit fit for HD\,188981B 
(top) and residuals to the fit (bottom).  The data are phase-folded and
two cycles are shown for clarity. }
 \label{binariesfits4}
\end{figure}

HD\,188981 is a known spectroscopic binary \citep{salzer85, sb9} that 
escaped the elimination of binaries in our initial target selection 
process.  Our iodine-cell velocity data have resulted in an extremely 
precise orbital solution for this system (Table~\ref{planetparams} and 
Figure~\ref{binariesfits4}).  We obtain m~sin~$i$=220.9\,\Mjup\ 
($=0.211$\,\Msun).  \citet{salzer85} remarked that the host star is 
$\sim$0.1 mag bluer than expected for a solitary K1 giant, and 
postulated that the stellar companion was an F-type dwarf.  An F dwarf 
companion of 1.2\,\Msun\ would require a binary orbital inclination of 
only $i\sim$10\degrees.  While the spectral lines from such a star would 
have remained hidden in the data of \citet{salzer85}, the higher 
resolution and S/N of the AAT data should have produced a double-lined 
spectrum, which would have caused the planet-search Doppler analysis to 
fail as for those stars in Table~\ref{sb2}.  \textbf{We simulated the 
effect of a second light at various levels of contamination.  These 
tests revealed that for a 10\% flux contribution from a secondary set of 
stellar lines, the Doppler code still produces reasonable though 
somewhat less precise velocities.  From Table~\ref{stellarparameters}, 
the luminosity of the primary is 12.3$\pm$0.9\,\Lsun, and hence a 10\% 
contamination from the secondary would give $L\ltsimeq$1.2\Lsun, 
corresponding to $\sim$1.05\,\Msun by the main-sequence mass-luminosity 
relation.  In other words, the lack of a detectable second set of 
stellar lines at best disfavours very high-mass companions. }


\section{Discussion and Conclusions}

Radial velocity observations of binaries are important in order to 
determine the masses, arguably the most fundamental parameter of stars.  
When combined with astrometric observations and parallax measurements, 
the true masses of the stars can be determined.  With the highly 
anticipated results from the GAIA mission \citep{perryman01}, it is 
important to have radial velocity observations of binary stars in hand 
as well.

\citet{rag10} presented a comprehensive study of stellar multiplicity 
for Sun-like stars in the solar neighbourhood, but their sample did not 
include evolved stars, and it remains unclear how the orbital parameters 
of a binary system are affected by a component's post-main-sequence 
evolution (but see Veras et al. 2011 and Mustill et al. 2012 for 
discussion of the effect on planets in such systems).  The stellar 
mass-ratio distribution prefers equal-mass pairs \citep{rag10}, but none 
of the four systems presented here has a mass ratio close to unity.  
However, this is a selection effect, as stars with a mass ratio close to 
unity would fall into the category of SB2s, which were dropped from this 
survey.  Because suspected binaries were initially avoided in the sample 
selection \citep{47205paper}, any newly discovered binaries will help 
estimate the number of binaries that have been missed in past 
estimations of the binary fraction \citep{dm91, rag10}, or verify their 
incompleteness corrections, respectively.  The aforementioned selection 
biases also make it difficult to compare the binary fraction for evolved 
stars to that of main sequence stars.

The planet HD\,155233b has been independently detected by the EXPRESS 
survey of M. Jones et al. (2015, in prep).  This planet joins the ranks 
of ``typical'' planets found to orbit intermediate-mass stars.  Such 
planets are characterised by super-Jupiter masses and semimajor axes 
beyond $\sim$1\,AU \citep{bowler10, johnson10b}.  \textbf{For 
intermediate-mass giants such as HD\,155233, giant planet occurrence has 
been shown to correlate positively with host-star mass 
\citep{johnson10}.  HD\,155233 is slightly metal-rich 
([Fe/H]=0.10$\pm$0.07), again consistent with the overall trend of 
increasing giant planet occurrence rate with host-star metallicity 
\citep{reffert15}.  }

\textit{Hipparcos} astrometry has been used for some massive companions 
to place limits on the system inclination, and hence the true masses 
\citep[e.g.][]{kurster08, reffert11, 91669paper}.  However, for the 
stars considered here, the distances are simply too large to make this 
approach useful at present.  Using the distances and the semimajor axes 
(Table~\ref{planetparams}), we estimate maximum angular separations as 
follows: HD\,34851B -- 0.044 mas; HD\,94386B -- 0.48 mas; HD\,104358B -- 
0.11 mas; HD\,155233b -- 0.48 mas; HD\,188981B -- 0.11 mas.  The greater 
astrometric precision of GAIA, of order 10 micro-arcsec for these bright 
targets \citep{perryman14}, will shed more light on these binary 
systems.  

Since the host stars are old, we are able to investigate whether our 
newly found large companions are detectable via direct imaging with the 
latest generation of instruments (GPI \& SPHERE) by assuming their 
luminosities to be similar to that of field main sequence dwarfs. 
Assuming that the system is edge on (i.e.~taking the minimum mass) we 
could use observed absolute magnitudes as a function of spectral type 
(collated and plotted in Kirkpatrick et al.~2012) to estimate the 
contrast ratio of each of our companions and their hosts.  By also 
assuming we are able to observe at the optimal time (maximum projected 
separation) we use the known distance to the systems to calculate their 
maximum on-sky separation.  We find that although the contrast ratios of 
these systems are not very large, the very small angular separations 
cause most of these systems to be inaccessible via direct imaging.  Only 
HD\,94386B is potentially detectable via observations using 
Non-Redundant Masking in conjunction with either SPHERE or GPI.  
Although HD\,155233b's projected separation is possibly feasible with 
Non-Redundant Masking, it is a giant planet in an old system, and 
therefore too faint to be detected with these types of observations 
(Table~\ref{contrasts}).


\section*{Acknowledgments}

We gratefully acknowledge the efforts of PPPS guest observers Brad 
Carter, Hugh Jones, and Simon O'Toole.  This research has made use of 
NASA's Astrophysics Data System (ADS), and the SIMBAD database, operated 
at CDS, Strasbourg, France.  This research has also made use of the 
Exoplanet Orbit Database and the Exoplanet Data Explorer at 
exoplanets.org \citep{wright11}.


\label{lastpage}


\begin{table}
  \centering
  \caption{Double-lined spectroscopic binaries in the PPPS sample}
  \begin{tabular}{ll}
  \hline
HD & HIP \\
\hline
749 & 944 \\
5873 & 4696 \\
5877 & 4618 \\
20035 & 14868 \\
31860 & 23061 \\
46122 & 31118 \\
58540 & 35790 \\
76321 & 43772 \\
81410 & 46159 \\
98579 & 55374 \\
136905 & 73525 \\
137164 & 75689 \\
142384 & 78027 \\
153438 & 83224 \\
176650 & 93383 \\
176794 & 94208 \\
204203 & 105953 \\
\hline
 \end{tabular}
\label{sb2}
\end{table}


\begin{table}
  \centering
  \caption{Summary of observations}
  \begin{tabular}{llll}
  \hline
Star & $N_{obs}$ & Span (days) & Mean uncertainty (\ms) \\
\hline
HD 34851   & 9 & 1278 & 6.0 \\ 
HD 47205   & 27 & 1881 & 1.1 \\
HD 94386   & 14 & 1512 & 1.8 \\
HD 104358  & 12 & 1880 & 2.7 \\
HD 155233  & 21 & 1426 & 2.1 \\
HD 188981  & 16 & 1453 & 2.3 \\
\hline
 \end{tabular}
\label{obslog}
\end{table}


\begin{table}
  \centering
  \caption{AAT radial velocities for HD\,34851}
  \begin{tabular}{lll}
  \hline
BJD-2400000 & Velocity (\ms) & Uncertainty (\ms) \\
 \hline
55251.94376  &    1418.0  &    3.6  \\
55525.13661  &  -15566.3  &    7.3  \\
55580.07072  &  -12641.7  &    6.8  \\
55602.00949  &     334.7  &    3.5  \\
55879.17147  &   -1270.4  &    6.7  \\
55906.06875  &   -8285.2  &    6.2  \\
55969.98892  &   -6009.1  &    6.8  \\
56375.89323  &       0.0  &    6.4  \\
56529.27818  &   -8020.4  &    6.4  \\ 
\hline
 \end{tabular}
\label{34851vels}
\end{table}

\clearpage


\begin{table}
  \centering
  \caption{AAT radial velocities for HD\,47205}
  \begin{tabular}{lll}
  \hline
BJD-2400000 & Velocity (\ms) & Uncertainty (\ms) \\
 \hline
54866.09965  &      28.1  &    0.8  \\
54866.94000  &      18.4  &    1.3  \\
54867.91576  &      26.6  &    1.3  \\
54869.08575  &      20.8  &    1.0  \\
54871.03478  &      30.4  &    1.3  \\
55140.18899  &     -29.1  &    0.9  \\
55227.06602  &     -18.4  &    1.3  \\
55317.85835  &       0.8  &    0.6  \\
55525.22369  &      34.4  &    1.4  \\
55526.21028  &      38.8  &    0.9  \\
55581.09317  &      43.3  &    1.1  \\
55601.00002  &      36.9  &    0.9  \\
55706.84304  &      -5.3  &    1.0  \\
55783.30462  &     -25.6  &    1.0  \\
55879.26442  &     -46.9  &    1.1  \\
55880.21953  &     -38.9  &    0.9  \\
55906.04456  &     -38.4  &    1.1  \\
55969.96686  &     -37.2  &    0.8  \\
55994.95994  &     -20.4  &    0.9  \\
56051.86418  &       1.6  &    1.4  \\
56059.86471  &      -4.0  &    1.6  \\
56343.99185  &      47.7  &    0.9  \\
56374.88203  &      55.7  &    1.2  \\
56377.97935  &      54.8  &    0.9  \\
56526.27125  &      -4.2  &    1.0  \\
56685.97593  &     -27.9  &    0.9  \\
56747.92127  &     -22.7  &    1.3  \\
\hline
 \end{tabular}
\label{47205vels}
\end{table}


\begin{table}
  \centering
  \caption{AAT radial velocities for HD\,94386}
  \begin{tabular}{lll}
  \hline
BJD-2400000 & Velocity (\ms) & Uncertainty (\ms) \\
 \hline
54866.16823  &   -1122.3  &    1.5  \\
55227.08417  &     386.9  &    1.2  \\
55380.83884  &    2145.4  &    1.2  \\
55580.16602  &     490.7  &    2.0  \\
55601.19789  &      15.7  &    1.8  \\
55602.15229  &      -4.4  &    1.6  \\
55706.90383  &   -1032.6  &    1.3  \\
55970.13115  &    -664.4  &    1.1  \\
55994.09656  &    -556.4  &    2.6  \\
56059.90490  &    -237.2  &    2.5  \\
56088.91084  &     -69.2  &    1.9  \\
56345.09758  &    2755.5  &    2.6  \\
56375.98219  &    3125.8  &    1.6  \\
56378.02772  &    3162.6  &    1.6  \\
\hline
 \end{tabular}
\label{94386vels}
\end{table}


\begin{table}
  \centering
  \caption{AAT radial velocities for HD\,104358}
  \begin{tabular}{lll}
  \hline
BJD-2400000 & Velocity (\ms) & Uncertainty (\ms) \\
 \hline
54867.22642  &   -2349.3  &    2.0  \\
55706.89202  &   -2436.0  &    1.9  \\
55757.87488  &    -745.0  &    5.4  \\
55970.17721  &   -2479.1  &    1.9  \\
55994.10810  &   -2265.1  &    4.2  \\
56059.94668  &     -66.3  &    2.9  \\
56086.95956  &     543.4  &    2.2  \\
56088.92816  &     596.7  &    2.2  \\
56344.12214  &       0.0  &    3.7  \\
56377.03081  &     683.3  &    2.4  \\
56378.03502  &     708.8  &    2.3  \\
56747.00059  &     609.4  &    2.0  \\
\hline
 \end{tabular}
\label{104358vels}
\end{table}

\clearpage

\begin{table}
  \centering
  \caption{AAT radial velocities for HD\,155233}
  \begin{tabular}{lll}
  \hline
BJD-2400000 & Velocity (\ms) & Uncertainty (\ms) \\
 \hline
55319.20751  &     -76.0  &    2.1  \\
55382.09355  &     -68.9  &    1.6  \\
55602.27053  &     -10.5  &    2.8  \\
55707.20272  &      12.8  &    1.8  \\
55757.95688  &       3.4  &    2.5  \\
55760.09094  &      24.8  &    2.0  \\
55841.92832  &      11.7  &    2.9  \\
55970.27601  &     -12.2  &    1.4  \\
55994.24764  &       9.4  &    2.1  \\
56052.14878  &     -48.3  &    2.4  \\
56089.07696  &     -16.1  &    1.8  \\
56134.99850  &     -29.0  &    1.9  \\
56344.27771  &     -12.7  &    1.6  \\
56375.25095  &      -0.2  &    2.4  \\
56376.23958  &      11.5  &    1.6  \\
56400.13392  &      -3.8  &    1.6  \\
56470.07735  &       2.9  &    3.8  \\
56494.98115  &      22.1  &    1.6  \\
56525.93237  &      37.0  &    1.8  \\
56529.98008  &      18.1  &    2.3  \\
56745.23184  &      47.2  &    1.2  \\
\hline
 \end{tabular}
\label{155233vels}
\end{table}


\begin{table}
  \centering
  \caption{AAT radial velocities for HD\,188981}
  \begin{tabular}{lll}
  \hline
BJD-2400000 & Velocity (\ms) & Uncertainty (\ms) \\
 \hline
55074.06452  &    -399.2  &    1.9  \\
55318.32115  &   -8926.5  &    2.5  \\
55455.97154  &    7190.8  &    1.9  \\
55670.31992  &       0.0  &    1.4  \\
55707.29018  &    6601.1  &    1.5  \\
55760.04921  &   -8824.9  &    6.1  \\
55842.90819  &    6847.6  &    2.0  \\
55994.29530  &   -3545.9  &    2.2  \\
56052.20653  &   -1600.1  &    1.7  \\
56060.24069  &   -4710.4  &    3.0  \\
56089.16329  &    8535.9  &    1.6  \\
56469.14010  &    8110.5  &    2.7  \\
56470.23672  &    7760.3  &    2.0  \\
56494.06543  &   -2017.2  &    1.7  \\
56495.07148  &   -2425.4  &    1.8  \\
56527.06483  &    7795.5  &    2.1  \\
\hline
 \end{tabular}
\label{188981vels}
\end{table}


\begin{table*}
  \centering
  \caption{Stellar Parameters for Host Stars}
  \begin{tabular}{l lc|lc|lc|lc|lc}
  \hline
                   &\multicolumn{2}{c}{HD 34851}               &\multicolumn{2}{c}{HD 94386}             &\multicolumn{2}{c}{HD 104358}          &\multicolumn{2}{c}{HD 155233}         &\multicolumn{2}{c}{HD 188981} \\
Parameter          & Value             & Ref.                  & Value          & Ref.                   & Value           & Ref.                & Value           & Ref.               & Value           & Ref.                 \\
 \hline
Spec.~Type         & K2 III            & 2                     & K3 IV          & 7                      & K0 III          & 8                   & K1 III          & 8                  & K1 III          & 10                   \\
                                                               & K2 III         & 8                                                                                                                                              \\
$(B-V)$            & 1.095             & 3                     & 1.176          & 3                      & 1.136           & 3                   & 1.030$\pm$0.008 & 4                  & 1.051           & 3                    \\
$E(B-V)$           & 0.0436            &                       & 0.0189         &                        & 0.0367          &                     & 0.0338          &                    & 0.0232          &                      \\
$A_V$              & 0.1359            &                       & 0.0590         &                        & 0.1146          &                     & 0.106           &                    & 0.0724          &                      \\
Mass (\Msun)       & 2.00$\pm$0.22     & 1                     & 1.19$\pm$0.19  & 1                      & 1.27$\pm$0.21   & 1                   & 1.50$\pm$0.20   & 1                  & 1.49$\pm$0.20   & 1                    \\
                                                               & 1.1            & 7                                                                                                     & 1.1             & 7                    \\
Distance (pc)      & 161.6$\pm$12.5    & 4                     & 73.8$\pm$4.7   & 4                      & 150.2$\pm$15.8  & 4                   & 75.1$\pm$3.3    & 4                  & 58.9$\pm$2.2    & 4                    \\
V sin $i$ (\kms)   & 4.6$\pm$0.9       & 1                     & $<$1           & 1                      & 4.1$\pm$0.5     & 1                   & 4.4$\pm$1.0     & 1                  & $<$1            & 1                    \\
                                                               & $<$1           & 7                                                                                                     & $<$1            & 7                    \\
$[Fe/H]$           & 0.29$\pm$0.14     & 1                     & 0.19$\pm$0.10  & 1                      & 0.06$\pm$0.08   & 1                   & 0.10$\pm$0.07   & 1                  & 0.18$\pm$0.07   & 1                    \\
                                                               & 0.08$\pm$0.09  & 7                      & 0.08$\pm$0.09   & 7                                                          & 0.08$\pm$0.09   & 7                    \\
T$_{\rm{eff}}$ (K)      & 4787$\pm$100      & 1                     & 4558$\pm$100   & 1                      & 4631$\pm$100    & 1                   & 4845$\pm$100    & 1                  & 4802$\pm$100    & 1                    \\
                   & 4815              & 5                     & 4436$\pm$13    & 9                      & 4400            & 5                   & 4436$\pm$13     & 9                  & 4436$\pm$13     & 9                    \\
                   & 4804$\pm$200      & 6                     & 4545$\pm$42    & 7                      & 4656$\pm$200    & 6                   & 4545$\pm$42     & 7                  & 4545$\pm$42     & 7                    \\
log $g$            & 3.05$\pm$0.09     & 1                     & 2.80$\pm$0.10  & 1                      & 2.80$\pm$0.12   & 1                   & 3.21$\pm$0.08   & 1                  & 3.20$\pm$0.08   & 1                    \\
                                                               & 2.7            & 9                                                              & 2.7             & 9                  & 2.7             & 9                    \\
                                                               & 2.7$\pm$0.3    & 7                                                              & 2.7$\pm$0.3     & 7                  & 2.7$\pm$0.3     & 7                    \\
Luminosity (\Lsun) & 23.1$\pm$3.6      & 1                     & 20.1$\pm$2.6   & 1                      & 22.7$\pm$4.8    & 1                   & 12.5$\pm$1.1    & 1                  & 12.3$\pm$0.9    & 1                    \\
Radius (\Rsun)     &  7.0$\pm$0.6      & 1                     & 7.2$\pm$0.6    & 1                      & 7.4$\pm$0.8     & 1                   & 5.03$\pm$0.22   & 1                  & 5.0$\pm$0.3     & 1                    \\
 \hline
 \multicolumn{11}{l}{Referneces: 1 - This work, 2 - \citet{houk75}, 3 - \citet{perryman97}, 4 - \citet{vl07}, 5 - \citet{mcdonald12},  }\\
 \multicolumn{11}{l}{6 - \citet{bailer11}, 7 - \citet{randich99}, 8 - \citet{houk88}, 9 - \citet{mass08}, 10 - \citet{houk82},  }\\
\end{tabular}
\label{stellarparameters}
\end{table*}

\clearpage


\begin{table*}
  \centering
  \caption{New and updated Keplerian orbital solutions }
  \begin{tabular}{lllllll}
  \hline
Parameter & HD 34851B & HD 47205 b & HD 94386 B & HD 104358 B & HD 
155233 b & HD 188981 B \\
 \hline
Period (days) & 62.304$\pm$0.005 & 796.0$\pm$7.4 & 925.0$\pm$1.0 & 
281.1$\pm$0.3 & 885$\pm$63 & 62.9637$\pm$0.0008 \\
$T_0$ (BJD-2400000) & 55214.7$\pm$0.4 & 54093$\pm$34 & 54582.1$\pm$1.2 & 
54836.8$\pm$4.2 & 55112$\pm$412 & 55011.487$\pm$0.008 \\ 
Eccentricity & 0.21$\pm$0.01 & 0.22$\pm$0.07 & 0.422$\pm$0.004 & 
0.24$\pm$0.02 & 0.03$\pm$0.20 & 0.4218$\pm$0.0007 \\
$\omega$ (degrees) & 203$\pm$2 & 77$\pm$14 & 37.8$\pm$0.2 & 146$\pm$7 & 
95$\pm$90 & 274.60$\pm$0.06 \\
$K$ (\ms) & 10296$\pm$144 & 41.8$\pm$2.4 & 2176$\pm$12 & 1828$\pm$42 & 
32.2$\pm$8.7 & 8735.6$\pm$2.2 \\ 
m sin $i$ (\Mjup) & 345.4$\pm$5.4 & 2.46$\pm$0.14 & 112.5$\pm$0.5 & 
69.3$\pm$1.9 & 2.0$\pm$0.5 & 220.90$\pm$0.08 \\
$a$ (AU) & 0.4078$\pm$0.0003 & 1.93$\pm$0.01 & 2.0266$\pm$0.0015 & 
0.9251$\pm$0.0007 & 2.07$\pm$0.10 & 0.369753$\pm$0.000005 \\
RMS about fit (\ms) & 34.6 & 6.5 & 5.7 & 15.3 & 10.9 & 5.3 \\
 \hline
 \end{tabular}
\label{planetparams}
\end{table*}


\clearpage


\clearpage


\clearpage


\begin{table*}
  \centering

  \caption{Estimated contrast ratios.  Companion masses are assumed to 
be the minimum m~sin~$i$ as given in Table~\ref{planetparams}, and 
separations are taken to be the maximum for each object modulo its 
orbital eccentricity and semimajor axis. }

  \begin{tabular}{llllllll}
  \hline
Host & SpT & Companion mass & SpT-companion & Separation & Distance 
 & Contrast & On-sky separation \\
   &   & (\Msun) &   & (AU) & (pc) & (mag) & (mas) \\
 \hline
HD\,34851B & K2 III & $>0.33$ & $<$M4 & $<0.49$ & 161.6 & $<3.9$ & $<3$ \\
HD\,94386B & K3 III & $>0.11$ & $<$M8/L0 & $<2.88$ & 73.8 & $<5.3$ & $<39$ \\
HD\,104358B & K0 III & $>0.066$ & $<$L5 & $<1.147$ & 150.2 & $<7.4$ & $<7.6$ \\
HD\,188981B & K1 III & $>0.211$ & $<$M6 & $<0.52$ & 58.9 & $<5.0$ & $<8.8$ \\
HD\,155233b & K1 III & $>2.13$ & $>$2.0\,\Mjup & $<2.13$ & 75.1 & ... & $<28.3$ \\
\hline
 \end{tabular}
\label{contrasts}
\end{table*}

\end{document}